# The use of percentiles and percentile rank classes in the analysis of bibliometric data: Opportunities and limits

Lutz Bornmann,* Loet Leydesdorff**, Rüdiger Mutz***

* First author and corresponding author:

Division for Science and Innovation Studies

Administrative Headquarters of the Max Planck Society

Hofgartenstr. 8,

80539 Munich, Germany.

E-mail: bornmann@gv.mpg.de

** Second author: Amsterdam School of Communication Research (ASCoR), University of

Amsterdam, Kloveniersburgwal 48, 1012 CX Amsterdam, The Netherlands;

loet@leydesdorff.net

*** Third author: Professorship for Social Psychology and Research on Higher Education
ETH Zurich (Swiss Federal Institute of Technology Zurich)
Muehlegasse 21
8001 Zurich, Switzerland
E-mail: mutz@gess.ethz.ch

**Abstract**

Percentiles have been established in bibliometrics as an important alternative to mean-based indicators for obtaining a normalized citation impact of publications. Percentiles have a number of advantages over standard bibliometric indicators used frequently: for example, their calculation is not based on the arithmetic mean which should not be used for skewed bibliometric data. This study describes the opportunities and limits and the advantages and disadvantages of using percentiles in bibliometrics. We also address problems in the calculation of percentiles and percentile rank classes for which there is not (yet) a satisfactory solution. It will be hard to compare the results of different percentile-based studies with each other unless it is clear that the studies were done with the same choices for percentile calculation and rank assignment.

# 1     Using reference sets

Since the 1980s, reference sets have been used for the normalization of citations in bibliometrics (see the overview in Vinkler, 2010). The purpose of reference sets is to allow the citation impact of a publication to be evaluated in comparison with the citation impact of similar publications. The use of reference sets is necessary as numerous studies have shown that citations depend not only on the quality of a publication, but also on many other factors. The factors which are commonly considered in bibliometrics to have a significant impact on citations are the subject category/area, the publication year and the document type of a publication. Many studies (Bornmann & Daniel, 2008) have revealed that other factors, such as the number of authors or the length of a publication, influence citations systematically; however, these other factors are not taken account of in the creation of reference sets. Presumably because it is not possible to assume a significant influence in all fields, and not all studies could consistently show a significant influence.

Since reference sets have been used, arithmetic means have been calculated to normalize citations of a publication or to make them comparable with others (see an overview on the different methods in Vinkler, 2012). The reference set was made up of publications which were published in the same subject area, the same publication year and with the same document type as the publications in question. The arithmetic mean value of the citations for the publications in the reference set was calculated to specify an expected citation impact (Schubert & Braun, 1986). With the citation impact of the publications in question and the expected citation impact a quotient was build: the mean observed citation rate (MOCR) divided by the mean expected citation rate (MECR). Using this quotient (the Relative Citation Rate, RCR) instead of raw citation counts, for example, the citation impact of a publication in the subject area of chemistry published five years ago was held to be comparable with the impact of a physics publication published ten years ago. Furthermore, it was now possible to analyze the overall citation impact for a publication set even if the publications were published in different fields and publication years or as different document types.

More recently, the use of RCR (or its derivatives such as CPP/FCSm, see Moed, Debruin, & Van Leeuwen, 1995) and NMCR (Glänzel, Thijs, Schubert, & Debackere, 2009) was criticized because a quotient does not allow for testing differences on their statistical significance (Opthof & Leydesdorff, 2010), respectively. The alternative measure (the mean normalized citation score, MNCS, Waltman, van Eck, van Leeuwen, Visser, & van Raan, 2011) is based on first normalizing each individual paper's citations against its reference standard and then taking the mean. However, there is a significant disadvantage inherent in the calculation of means for the normalization of citations (Bornmann, Mutz, Marx, Schier, & Daniel, 2011): As a rule, distribution of citation counts over publications sets is skewed. The arithmetic mean value calculated for a reference set may therefore be determined by a few highly cited publications. The arithmetic mean as a measure of central tendency is not suitable for skewed data (Bornmann, et al., 2011).

For example, in the Leiden Ranking 2011/2012 the University of Göttingen is only in position 2 in a ranking based on mean-based quotients (MNCS) because the indicator for this university "turns out to have been strongly influenced by a single extremely highly cited publication" (Waltman et al., 2012, p. 10). Attributions which would thus describe the citation impact as excellent or outstanding are based on erroneous assumptions about the citation distributions. Given the skewedness of the distributions, nonparametric statistics should be used (Leydesdorff, Bornmann, Mutz, & Opthof, 2011).

## 2    The calculation of percentiles

A percentile is a value below which a certain proportion of observations fall. Percentiles (and therefore also percentile rank classes, PRs) offer an alternative to mean-based quotients. Generally speaking, (1) percentiles are not as strongly influenced by extreme values of the distribution as the mean value (Waltman, et al., 2012, p. 10); (2) they do not depend on the choice of a specific probability density function in comparison to the arithmetic mean, which requires normally distributed data. (3) Percentiles can be calculated, even if the data are skewed, as it is usual for bibliometric data. (4) Once calculated, they also permit the classification of publications into meaningful citation impact classes (Bornmann, in press-b; Leydesdorff, in press; Rousseau, 2012; Schreiber, 2012). With a reference set it is possible to determine, for example, whether a publication belongs to the 1% or 10% of the most cited publications in the reference set or not.

Three steps are needed in order to calculate the percentiles for a reference set, (see sample data in Table 1):

First, the rank-frequency function (see Egghe & Rousseau, 2006) is calculated. All publications in the set are ranked in decreasing order by their number of citations (see column i in Table 1), and the number of publications in the (reference) set is determined (see column n). Publications with equal citation counts are assigned the average rank. For example, 41 publications are shown in the table. As there are two publications with the highest number of citations (15), they are assigned the rank 40.5 instead of ranks 41 and 40. This is the default ranking method in StataCorp. (2011) which ensures that the sum of the ranks is preserved. Schreiber (2012) proposed to average the percentile values and not the rank values.

Secondly, the minimum or maximum, respectively, of the percentile scale must be determined. Publications with 0 citations should be assigned a percentile of zero (see Percentile column). This rule is applied for example in a modified form by InCites (Thomson Reuters), as publications with 0 citations are always assigned the value 100 (or 1) in this database. InCites (http://incites.thomsonreuters.com/) is a web-based research evaluation tool allowing assessment of the productivity and citation impact of institutions. In InCites, publications with a high citation impact are assigned a low percentile and publications with a low citation impact are assigned a high percentile. Thus, percentiles from InCites will be named in the following as "inverted percentiles." It is only by assigning the value 0 (or 100 in InCites) to the publications with 0 citations that it is ensured that the missing citation impact of publications is reflected in the percentiles in the same way in every case. Different values for publications with 0 citations would arise if percentiles are calculated without using a constant value of zero.

Thirdly, each publication is assigned a percentile based on the citation distribution (sorted in decreasing order). However, percentiles can be calculated in different ways (Cox, 2005). In Table 1, for example, the publication halfway down the ranking is the 21$^{st}$ publication with 4 citations. This publication (the median) should be assigned the percentile 50 (or 0.5) as 50% of the publications in the table have a higher rank (more citations) and 50% of the values have a lower rank (fewer citations).

If one were to create the quotients using the ranks (i) and the number of publications (n) (i/n*100), in order to calculate the quantiles – that is, the continuous variable from which percentiles can be derived by rounding – this would result in the quotients 0 (for 0 citations, see above), 19.51 ... 98.78 (see column percentile (a)). This calculation is used for the percentile data in InCites (with the exception of assigning average ranks to equal citation counts, see the first step above) and, for example, by Rousseau (2012). However, it is neither the case that (21/41*100) results in the mean percentile of 50 (but in 51.22) nor does this value handle the tails symmetrically (50% of the other publications should be below and 50% above the publication with four citations). If instead, one uses the formula ((i-1)/n*100) for the calculation to get a mean percentile of 50 (see column percentile (b)) the following percentiles would result: 0 (for 0 citations), 17.07... 96.34. Again, this would not result in a mean percentile of 50 but 48.78, nor in a symmetrical treatment of the tails.

The compromise ((i-0.5)/n*100) derived by (Hazen, 1914) is used very frequently nowadays for the calculation of percentiles (for example by StataCorp., 2011). Leydesdorff (in press) specifies the uncertainty in the percentile attribution as 0.5*n. Using (i/n), this correction leads to (i/n)-(0.5/n)=(i-0.5)/n in the case of using Rousseau's (2012) rules for the computation or ((i-1)/n)+(0.5/n)=(i-0.5)/n using Leydesdorff et al.'s (2011) rules for the computation. In both cases, the result is the same as Hazen's (1914) compromise value. As the column percentile (c) in the table shows, this method results in the desired value of 50 for the publication with 4 citations.

Many other proposals for the calculation of percentiles made under the label "plotting position problem" following Hazen (1914), can be summarized using the following formula: (i-a)/(n-2a+1) (Cunnane, 1978; Harter, 1984). In the proposal by Hazen (1914) a=0.5. Blom (1958) proposes, for example, a=0.375, particularly when the values are distributed normally in a set (see column percentile (d) in the table). With a=0.375 the middle publication in the table has a percentile of 50.31 not 50; however the distribution of the citations is not normal, but skewed. For the exponential distribution, Gringorten (1963) recommended the following formula for the calculation of percentiles: (i-0.44)/(n+0.12)*100 (see column percentile (e) in the table). Similarly to the proposal by Hazen (1914) this also results in a value of 50 for the mean value in the table. This formula would therefore also be suitable for the percentile calculation for the publication set.

It is already possible to achieve meaningful results from an analysis of percentiles. For example, Figure 1 shows violin plots (Hintze & Nelson, 1998) for four different universities. Each publication by the four universities is assigned to a percentile corresponding to its reference set (after normalization for the respective subject category, publication year, and document type). The inverted percentiles are derived from InCites, which is why in Figure 1 a higher citation impact of the publications is associated with decreasing values. The violin plots show a marker for the median of the inverted percentiles, a box indicating the interquartile range, and spikes extending to the upper and lower-adjacent values. The upper-adjacent value is the highest value of the inverted percentiles which is smaller than or equal to the third quartile plus 1.5 times the interquartile range (Kohler & Kreuter, 2009) and amounts in our examples to zero citations. Furthermore, plots of the estimated kernel density (Hintze & Nelson, 1998) are overlaid. The figure clearly shows a significantly better performance for Univ 3 compared to the others. On the one hand, the median inverted percentile is lower than for the other universities. Furthermore, the distribution of this university is characterized by fewer publications with 0 citations (i.e. inverted percentile of 100) compared to the others. Figure 2 provides an alternative representation of the inverted percentiles for six universities. These are box plots which show the mean, the standard deviations and the range of the inverted percentiles for the universities.

## 3      Assigning percentiles to percentile rank classes

It is not only possible to calculate percentiles for the publications (of a university or a scientist and so on) but also to classify and evaluate them in percentile rank classes (PRs). Each individual publication can be assigned to a class using the relevant reference sets or the reference set is used to create a percentile for each publication which is then assigned to a PR. By assigning the percentiles to classes, it is possible to interpret the performance of a publication unambiguously. For example, scientist 1 has published 10 publications, of which half belong to the 10% most cited publication in the corresponding reference set; the number of publications in these classes can be compared with each other for different publication sets. If scientist 2 has published a third of her publications which belong to the 10% most cited publications, she has a better performance than scientist 1. Since the expected value for the 10% most cited publications is 10%, a scientist's percentage share of these publications in one class (for example, half the publications) can be compared with the expected value. If a sufficiently large publication set (for a scientist, a university etc.) were randomly selected from the percentiles for all the publications in a literature database (e.g. Web of Science, Thomson Reuters), one could expect that 10% of the publications belong to the 10% most cited publications in their reference sets.

Today, four different class schemes are common when categorizing percentiles. The different schemes are conventions for categorizing citation impact which have proven to be useful in practice but which cannot be justified scientifically (Leydesdorff, et al., 2011). (i) With PR(2, 10) the percentiles are categorized in the following two rank classes, where the first number denotes the number of classes (i.e., 2), the second number indicates the p% percentile (10% percentile). As it has now established as standard on an institutional level to designate those publications in a set which are among the 10% most cited publications in their subject area (publication year, document type) as highly cited publications (Bornmann, de Moya Anegón, & Leydesdorff, 2012; Bornmann & Leydesdorff, 2011; Tijssen & van Leeuwen, 2006; Tijssen, Visser, & van Leeuwen, 2002; Waltman, et al., 2012), the following two rank classes are formed:

(1) <90% (papers with a percentile smaller than the $90^{th}$ percentile),

(2) 10% (papers with a percentile equal to or larger than the $90^{th}$ percentile).

(ii) With PR(2, 50) the publications are also categorized into two rank classes but those publications which are either equal to or above the median (papers with a percentile equal to or larger than the $50^{th}$ percentile) or less than the median (papers with a percentile smaller than the $50^{th}$ percentile) in their reference set are classified into two classes.

(iii) With PR(6) the percentiles are categorized into six rank classes (Bornmann & Mutz, 2011). This approach is in use as the evaluation scheme in the Science and Engineering Indicators of the US National Science Foundation (National Science Board, 2012) since a number of years. In this scheme the focus is on publications cited more frequently than the median. The six percentile rank classes are aggregated as follows:

(1) <50% (papers with a percentile smaller than the $50^{th}$ percentile),

(2) 50% (papers within the [$50^{th}$; $75^{th}$[ percentile interval),

(3) 25% (papers within the [$75^{th}$; $90^{th}$[ percentile interval),

(4) 10% (papers within the [$90^{th}$; $95^{th}$[ percentile interval),

(5) 5% (papers within the [95$^{th}$; 99$^{th}$[ percentile interval),

(6) 1% (papers with a percentile equal to or larger than the 99$_{th}$ percentile).

(iv) The table "Percentiles for papers published by field" is offered by Thomson Reuters' Essential Science Indicators (ESI). The table in the ESI gives citation thresholds for 22 broad fields of this database (such as Chemistry or Clinical Medicine) which allow inclusion in the following six percentile rank classes:

(1) 50% (papers with a percentile equal to or larger than the 50$^{th}$ percentile).

(2) 20% (papers with a percentile equal to or larger than the 80$^{th}$ percentile).

(3) 10% (papers with a percentile equal to or larger than the 90$^{th}$ percentile).

(4) 1% (papers with a percentile equal to or larger than the 99$^{th}$ percentile).

(5) 0.1% (papers with a percentile equal to or larger than the 99.9$^{th}$ percentile).

(6) 0.01% (papers with a percentile equal to or larger than the 99.99$^{th}$ percentile).

The scheme used by Thomson Reuters gives a better indication of skew and ranking among the top-cited publications. If a user of the ESI knows the citations and the broad field for a publication he or she can use the citation thresholds in the percentile table to determine in which rank class the publication falls which gives an evaluation of the performance of a publication measured against the reference publications (note that ESI does not normalize for document types).

The PR can be used to provide a meaningful analysis of the citation impact of various publication sets. Figure 3 shows the differences in the various classes between four universities, based on PR(6) (see iii above) (Bornmann, in press-a). It shows the percentage share of publications from each university in the individual classes. For example, 43.38% of the publications from Univ 4 belong in class <50% and 1.55% in class 1% in their subject area. Univ 3 differs significantly from the other universities. For this university there are considerably fewer publications in the class <50% (31%) and (many) more publications in the class 5% (9.31%) and in the class 1% (2.89%) than for the other universities. The comparably good performance of Univ 3 is therefore traceable to the small proportion of little cited and the high proportion of much cited works.

Using the percentage values in Figure 3 it is not only possible to compare the universities with each other, but also to compare the percentage values of each university with expected values which would result from a random selection of publications from a database. For example, one can expect for each university that 1% of their publications belong in the class 1% and 4% in the class 5%. Thus, for example, with 9.31% (Univ 3), 5.55% (Univ 4), 7.08% (Univ 5) and 5.07% (Univ 6) all four universities published (significantly) more class 5% papers than the expected value of 4%. For the class 50%, for all four universities there are percentage values which roughly agree with the expected value of 25% (between 23.17% and 25.02%). The expected values form unambiguous reference values for the percentage values in the PR(6) which can be interpreted as a difference (deviation) from the reference value.

## 4   Rules for and problems with assigning publications to percentile rank classes

These two rules should be complied with when PRs are created for the publications in a reference set. Whereas the rule under 1 refers to the number of citations, rule under 2 focuses on the number of publications:

(1) The uncertain assignment of publications to PRs causes problems if there are more publications with the same citation counts in a reference set than would be assigned to a class. For example, if, as in Table 1, out of 41 publications there are 7 with 1 citation then it will not be possible to assign the publications to more than five classes with the same class sizes unequivocally. This means that quintiles are still a possibility, but sextiles are not.

(2) The number of classes should be less than or equal to the number of publications plus one. For example, this rule is used by StataCorp. (2011). If unequal class sizes are used, such as the PR(6) class 1%, class 5% and so on (see iii above), then the smallest of these class sizes should determine the number of classes chosen. Therefore, to use this PR(6), there should be at least 101 papers without ties in one reference set, as with the 1% class it is necessary to categorize 1% of the papers (i.e. classification in the 100 rank class scheme, PR(100)).

If, as is frequently the case in bibliometrics, the Web of Science subject categories are used to compile a reference set, it is reasonable to assume the possibility of creating up to 100 PRs. The number of publications of one document type published in a certain subject category and in a certain publication year is as a rule more than 101. However, it is also common in bibliometrics to use a single journal and not a subject category as a reference set (Vinkler, 2010). In this case, there is a risk (particularly with journals which predominantly publish reviews) that the case numbers are not sufficiently large to create a certain class.

Even if there is a sufficiently large number of publications in a reference set with which to create PRs, there can nevertheless be a problem with the unambiguous assignment of publications to classes. As there are 2 publications with 15 citations and 4 publications with 14 citations in Table 1, it is not possible to assign 10% of the publications unambiguously to the class 10% due to the tied citations. If one were to be guided by the classification of publications in the classes in the percentile (c) column in the table, one would classify 15% of the publications in the class 10% (all publications with a percentile >=90). This assignment problem depends primarily on the disparity (lack of equality) of the citations in a reference set. The more frequently certain citation counts occur more than once, the more difficult the classification. The disparity of the citations is dependent on the subject category, the length of the citation window, and the number of publications. One is more likely to find a greater disparity between citations in larger than in smaller reference sets and in sets used for a longer rather than a smaller citation window. Furthermore, the disparity in subject categories in the life sciences will be higher than in the physical sciences.

There are various options for dealing with the problem of insufficient citation disparity for creating PRs and the resulting uncertainty in the assignment of publications to classes. (i) Today it is (still) standard to accept this lack of clarity and to assign the publications to either the higher or the lower class. For the publications in Table 1, for example, this would mean that the publications with 15 citations (2 publications or 5% of the publications) or the six publications with 15 or 14 citations (15% of the publications) would all be assigned to the class 10%. (ii) Publications which cannot be assigned unambiguously to a class can be handled as missing values. However, it can happen as a result that a publication in question, for which the reference set was put together for its normalization, cannot be assigned a PR because it receives a missing value. (iii) There is another option of introducing a further publication attribute (in addition to the subject area, the document type and the publication year) as a covariate into the analysis and dividing the number of citations, for example, by the number of pages of a publication. As described above, a number of studies have shown that longer publications can be expected to have a higher citation impact than shorter publications. The resulting quotients should be more clearly rankable for the percentile calculation than the citation counts. A similar approach is used by the SCImago group for their institutions ranking (http://www.scimagoir.com). Where citation counts are equal, the SJR2 (Guerrero-Bote & de Moya-Anegon, 2012) of the journal which has published the publications is used as a second sort key (from highest to lowest). This journal metric takes into account not only the prestige of the citing scientific publication but also its closeness to the cited journal.

(iv) Waltman and Schreiber (2012) have proposed a fourth option: Publications which cannot be assigned unambiguously to a PR are assigned on a fractional basis. For example, in Table 1 using PR(2,10) the four publications with 14 citations would be assigned to the class 10% with a weighting of 0.525 and to the class <90% with a weighting of 0.525. Adding 0.525*4 = 2.1 to the value of 2 for the two most-cited publications in the table, the result is 4.1 which is exactly 10% of 41. Following the proposal by Waltman and Schreiber (2012) ensures that one always obtains a proportion of 10% for the class 10% publications and proportion of 90% for the class <90% publications with one reference set.

The disadvantage of this approach is that there is no longer an unambiguous class assignment for the fractionally assigned publications. For example, Figure 3 could not easily be created on the basis of fractionally assigned publications because for some publications there would be no unambiguous classification. Using the weightings which the individual publications are given in the Waltman and Schreiber (2012) approach, it would however be possible to calculate the total publications in each PR. For example, if the publications from one university were assigned to the PR(2,10) it would be possible to determine the proportion of 10% class publications as is done in the Leiden Rankings 2011 (Waltman et al., in press). The resulting proportions of top-10% most highly cited papers can also be tested on their statistical significance (Leydesdorff & Bornmann, 2012).

Even though it is possible to determine the more precise proportion of publications in certain classes with the Waltman and Schreiber (2012) approach, the data can no longer be used for other statistical analyses at the level of the papers under study. At the paper level the uncertainty cannot be specified. For example, let us take PR(2,10) again. Say Bornmann (in press-a) calculates a multiple logistic regression model to identify citation impact differences between four universities. This model is appropriate for the analysis of binary responses. His binary response is coded 0 for a publication in the class 90% and 1 for a publication in the class 10%. As binary responses arise only when the outcome is the presence or absence of an event, a response variable, created in accordance with the Waltman and Schreiber (2012) approach can no longer be used as input for the logistic regression model. A different analysis method would have to be used which does not perform the analysis at the level of single publications.

# 5      Discussion

Percentiles and PR have been established in bibliometrics as an important alternative to the mean-based indicators with which to arrive at a normalized citation impact of publications. This study addresses some of the problems in calculating percentiles and PRs for which there is not (yet) a satisfactory solution.

(1) Percentiles can be calculated in different ways for a reference set. Nowadays, the Hazen (1914) formula is used very frequently. However, in bibliometrics it would be possible to follow the proposal in Gringorten (1963). More studies are needed here to look at the possible advantages and disadvantages of the various approaches to bibliometric data.

(2) We have shown in this study that PRs are very useful when used in evaluation studies to reveal the differences in performance between two research institutions, such as universities. However, the size of the reference set should always be taken into account when setting up the classes: The number of classes should always be less than or equal to the number of papers plus one.

There is a specific problem with the assignment of publications to PRs where the publications have the same number of citations. Some ways of dealing with this problem have been presented in this study. However, none of them is yet completely satisfactory and further investigation in future studies is therefore required. A procedure is sought which will allow every publication (as far as possible) with unambiguous citation thresholds to be assigned to meaningful PRs.

One has to be careful when applying percentile techniques and it will be hard to compare the results of different studies with each other unless it is clear that the studies were done with the same choices for percentile calculation and rank assignment.

# Acknowledgements

We would like to thank Ludo Waltman and Michael Schreiber for the stimulating discussion preceding this publication.

# References

Blom, G. (1958). *Statistical estimates and transformed beta-variables*. New York, NY, USA: John Wiley & Sons.

Bornmann, L. (in press-a). How to analyze percentile impact data meaningfully in bibliometrics? The statistical analysis of distributions, percentile rank classes and top-cited papers. *Journal of the American Society for Information Science and Technology*.

Bornmann, L. (in press-b). The problem of percentile rank scores used with small reference sets. *Journal of the American Society of Information Science and Technology*.

Bornmann, L., & Daniel, H.-D. (2008). What do citation counts measure? A review of studies on citing behavior. *Journal of Documentation, 64*(1), 45-80. doi: 10.1108/00220410810844150.

Bornmann, L., de Moya Anegón, F., & Leydesdorff, L. (2012). The new Excellence Indicator in the World Report of the SCImago Institutions Rankings 2011. *Journal of Informetrics, 6*(2), 333-335. doi: 10.1016/j.joi.2011.11.006.

Bornmann, L., & Leydesdorff, L. (2011). Which cities produce more excellent papers than can be expected? A new mapping approach—using Google Maps—based on statistical significance testing. *Journal of the American Society of Information Science and Technology, 62*(10), 1954-1962.

Bornmann, L., & Mutz, R. (2011). Further steps towards an ideal method of measuring citation performance: the avoidance of citation (ratio) averages in field-normalization. *Journal of Informetrics, 5*(1), 228-230.

Bornmann, L., Mutz, R., Marx, W., Schier, H., & Daniel, H.-D. (2011). A multilevel modelling approach to investigating the predictive validity of editorial decisions: do the editors of a high-profile journal select manuscripts that are highly cited after publication? *Journal of the Royal Statistical Society - Series A (Statistics in Society), 174*(4), 857-879. doi: 10.1111/j.1467-985X.2011.00689.x.

Cox, N. J. (2005). Calculating percentile ranks or plotting positions. Retrieved May 30, from http://www.stata.com/support/faqs/stat/pcrank.html

Cunnane, C. (1978). Unbiased plotting positions — A review. *Journal of Hydrology, 37*(3–4), 205-222. doi: 10.1016/0022-1694(78)90017-3.

Egghe, L., & Rousseau, R. (2006). An informetric model for the Hirsch-index. *Scientometrics, 69*(1), 121-129.

Glänzel, W., Thijs, B., Schubert, A., & Debackere, K. (2009). Subfield-specific normalized relative indicators and a new generation of relational charts: methodological foundations illustrated on the assessment of institutional research performance. *Scientometrics, 78*(1), 165-188.

Gringorten, I. I. (1963). A plotting rule for extreme probability paper. *Journal of Geophysical Research, 68*(3), 813-&.

Guerrero-Bote, V. P., & de Moya-Anegon, F. (2012). A further step forward in measuring journals' scientific prestige: the SJR2 indicator. Retrieved July 3, from http://arxiv.org/abs/1201.4639

Harter, H. L. (1984). Another look at plotting positions. *Communications in Statistics-Theory and Methods, 13*(13), 1613-1633.

Hazen, A. (1914). Storage to be provided in impounding reservoirs for municipal water supply. *Transactions of American Society of Civil Engineers, 77*, 1539-1640.

Hintze, J. L., & Nelson, R. D. (1998). Violin plots: a box plot-density trace synergism. *The American Statistician, 52*(2), 181-184.

Kohler, U., & Kreuter, F. (2009). *Data analysis using Stata* (2. ed.). College Station, TX, USA: Stata Press, Stata Corporation.

Leydesdorff, L. (in press). Accounting for the uncertainty in the evaluation of percentile ranks. *Journal of the American Society for Information Science and Technology*.

Leydesdorff, L., & Bornmann, L. (2012). Testing differences statistically with the Leiden ranking. *Scientometrics, 92*(3), 781-783.

Leydesdorff, L., Bornmann, L., Mutz, R., & Opthof, T. (2011). Turning the tables in citation analysis one more time: principles for comparing sets of documents. *Journal of the American Society for Information Science and Technology, 62*(7), 1370-1381.

Moed, H. F., Debruin, R. E., & Van Leeuwen, T. N. (1995). New bibliometric tools for the assessment of national research performance - database description, overview of indicators and first applications. *Scientometrics, 33*(3), 381-422.

National Science Board. (2012). *Science and engineering indicators 2012*. Arlington, VA, USA: National Science Foundation (NSB 12-01).

Opthof, T., & Leydesdorff, L. (2010). Caveats for the journal and field normalizations in the CWTS ("Leiden") evaluations of research performance. *Journal of Informetrics, 4*(3), 423-430.

Rousseau, R. (2012). Basic properties of both percentile rank scores and the I3 indicator. *Journal of the American Society for Information Science and Technology, 63*(2), 416-420. doi: 10.1002/asi.21684.

Schreiber, M. (2012). Inconsistencies of recently proposed citation impact indicators and how to avoid them. Retrieved February 20, from http://arxiv.org/abs/1202.3861

Schubert, A., & Braun, T. (1986). Relative indicators and relational charts for comparative assessment of publication output and citation impact. *Scientometrics, 9*(5-6), 281-291.

StataCorp. (2011). *Stata statistical software: release 12*. College Station, TX, USA: Stata Corporation.

Tijssen, R., & van Leeuwen, T. (2006). Centres of research excellence and science indicators. Can 'excellence' be captured in numbers? In W. Glänzel (Ed.), *Ninth International Conference on Science and Technology Indicators* (pp. 146-147). Leuven, Belgium: Katholieke Universiteit Leuven.

Tijssen, R., Visser, M., & van Leeuwen, T. (2002). Benchmarking international scientific excellence: are highly cited research papers an appropriate frame of reference? *Scientometrics, 54*(3), 381-397.

Vinkler, P. (2010). *The evaluation of research by scientometric indicators*. Oxford, UK: Chandos Publishing.

Vinkler, P. (2012). The case of scientometricians with the "absolute relative" impact indicator. *Journal of Informetrics, 6*(2), 254-264. doi: 10.1016/j.joi.2011.12.004.

Waltman, L., Calero-Medina, C., Kosten, J., Noyons, E. C. M., Tijssen, R. J. W., van Eck, N. J., . . . Wouters, P. (2012). The Leiden Ranking 2011/2012: data collection, indicators, and interpretation. Retrieved February 24, from http://arxiv.org/abs/1202.3941

Waltman, L., Calero-Medina, C., Kosten, J., Noyons, E. C. M., Tijssen, R. J. W., van Eck, N. J., . . . Wouters, P. (in press). The Leiden Ranking 2011/2012: data collection, indicators, and interpretation. *Journal of the American Society for Information Science and Technology*.

Waltman, L., & Schreiber, M. (2012). On the calculation of percentile-based bibliometric indicators. Retrieved May 4, from http://arxiv.org/abs/1205.0646

Waltman, L., van Eck, N. J., van Leeuwen, T. N., Visser, M. S., & van Raan, A. F. J. (2011). Towards a new crown indicator: some theoretical considerations. *Journal of Informetrics, 5*(1), 37-47. doi: 10.1016/j.joi.2010.08.001.

Table 1. The calculation of percentiles for a sample set of 41 publications

| No | Citations | i | n | percentile (a) | percentile (b) | percentile (c) | percentile (d) | percentile (e) |
|---|---|---|---|---|---|---|---|---|
| 41 | 15 | 40.5 | 41 | 98.78049 | 96.34146 | 97.56097 | 97.86585 | 97.42218 |
| 40 | 15 | 40.5 | 41 | 98.78049 | 96.34146 | 97.56097 | 97.86585 | 97.42218 |
| 39 | 14 | 37.5 | 41 | 91.46342 | 89.02439 | 90.2439 | 90.54878 | 90.12646 |
| 38 | 14 | 37.5 | 41 | 91.46342 | 89.02439 | 90.2439 | 90.54878 | 90.12646 |
| 37 | 14 | 37.5 | 41 | 91.46342 | 89.02439 | 90.2439 | 90.54878 | 90.12646 |
| 36 | 14 | 37.5 | 41 | 91.46342 | 89.02439 | 90.2439 | 90.54878 | 90.12646 |
| 35 | 12 | 33.5 | 41 | 81.70731 | 79.2683 | 80.48781 | 80.79269 | 80.39883 |
| 34 | 12 | 33.5 | 41 | 81.70731 | 79.2683 | 80.48781 | 80.79269 | 80.39883 |
| 33 | 12 | 33.5 | 41 | 81.70731 | 79.2683 | 80.48781 | 80.79269 | 80.39883 |
| 32 | 12 | 33.5 | 41 | 81.70731 | 79.2683 | 80.48781 | 80.79269 | 80.39883 |
| 31 | 10 | 30 | 41 | 73.17073 | 70.7317 | 71.95122 | 72.2561 | 71.88716 |
| 30 | 10 | 30 | 41 | 73.17073 | 70.7317 | 71.95122 | 72.2561 | 71.88716 |
| 29 | 10 | 30 | 41 | 73.17073 | 70.7317 | 71.95122 | 72.2561 | 71.88716 |
| 28 | 9 | 28 | 41 | 68.29269 | 65.85366 | 67.07317 | 67.37805 | 67.02335 |
| 27 | 8 | 27 | 41 | 65.85366 | 63.41463 | 64.63415 | 64.93903 | 64.59144 |
| 26 | 7 | 26 | 41 | 63.41463 | 60.97561 | 62.19512 | 62.5 | 62.15953 |
| 25 | 6 | 25 | 41 | 60.97561 | 58.53659 | 59.7561 | 60.06097 | 59.72763 |
| 24 | 5 | 23 | 41 | 56.09756 | 53.65854 | 54.87805 | 55.18293 | 54.86381 |
| 23 | 5 | 23 | 41 | 56.09756 | 53.65854 | 54.87805 | 55.18293 | 54.86381 |
| 22 | 5 | 23 | 41 | 56.09756 | 53.65854 | 54.87805 | 55.18293 | 54.86381 |
| 21 | 4 | 21 | 41 | 51.21951 | 48.78049 | 50 | 50.30488 | 50 |
| 20 | 3 | 18 | 41 | 43.90244 | 41.46341 | 42.68293 | 42.9878 | 42.70428 |
| 19 | 3 | 18 | 41 | 43.90244 | 41.46341 | 42.68293 | 42.9878 | 42.70428 |
| 18 | 3 | 18 | 41 | 43.90244 | 41.46341 | 42.68293 | 42.9878 | 42.70428 |
| 17 | 3 | 18 | 41 | 43.90244 | 41.46341 | 42.68293 | 42.9878 | 42.70428 |
| 16 | 3 | 18 | 41 | 43.90244 | 41.46341 | 42.68293 | 42.9878 | 42.70428 |
| 15 | 2 | 13.5 | 41 | 32.92683 | 30.4878 | 31.70732 | 32.0122 | 31.7607 |
| 14 | 2 | 13.5 | 41 | 32.92683 | 30.4878 | 31.70732 | 32.0122 | 31.7607 |
| 13 | 2 | 13.5 | 41 | 32.92683 | 30.4878 | 31.70732 | 32.0122 | 31.7607 |
| 12 | 2 | 13.5 | 41 | 32.92683 | 30.4878 | 31.70732 | 32.0122 | 31.7607 |
| 11 | 1 | 8 | 41 | 19.5122 | 17.07317 | 18.29268 | 18.59756 | 18.38521 |
| 10 | 1 | 8 | 41 | 19.5122 | 17.07317 | 18.29268 | 18.59756 | 18.38521 |
| 9 | 1 | 8 | 41 | 19.5122 | 17.07317 | 18.29268 | 18.59756 | 18.38521 |
| 8 | 1 | 8 | 41 | 19.5122 | 17.07317 | 18.29268 | 18.59756 | 18.38521 |
| 7 | 1 | 8 | 41 | 19.5122 | 17.07317 | 18.29268 | 18.59756 | 18.38521 |
| 6 | 1 | 8 | 41 | 19.5122 | 17.07317 | 18.29268 | 18.59756 | 18.38521 |
| 5 | 1 | 8 | 41 | 19.5122 | 17.07317 | 18.29268 | 18.59756 | 18.38521 |
| 4 | 0 | 2.5 | 41 | 0 | 0 | 0 | 0 | 0 |
| 3 | 0 | 2.5 | 41 | 0 | 0 | 0 | 0 | 0 |
| 2 | 0 | 2.5 | 41 | 0 | 0 | 0 | 0 | 0 |
| 1 | 0 | 2.5 | 41 | 0 | 0 | 0 | 0 | 0 |

Notes. no=publication number; citations=citation counts of the publication; i=rank of the publication in the set; n=number of publications; percentile (a) to percentile (e)=differently calculated percentiles

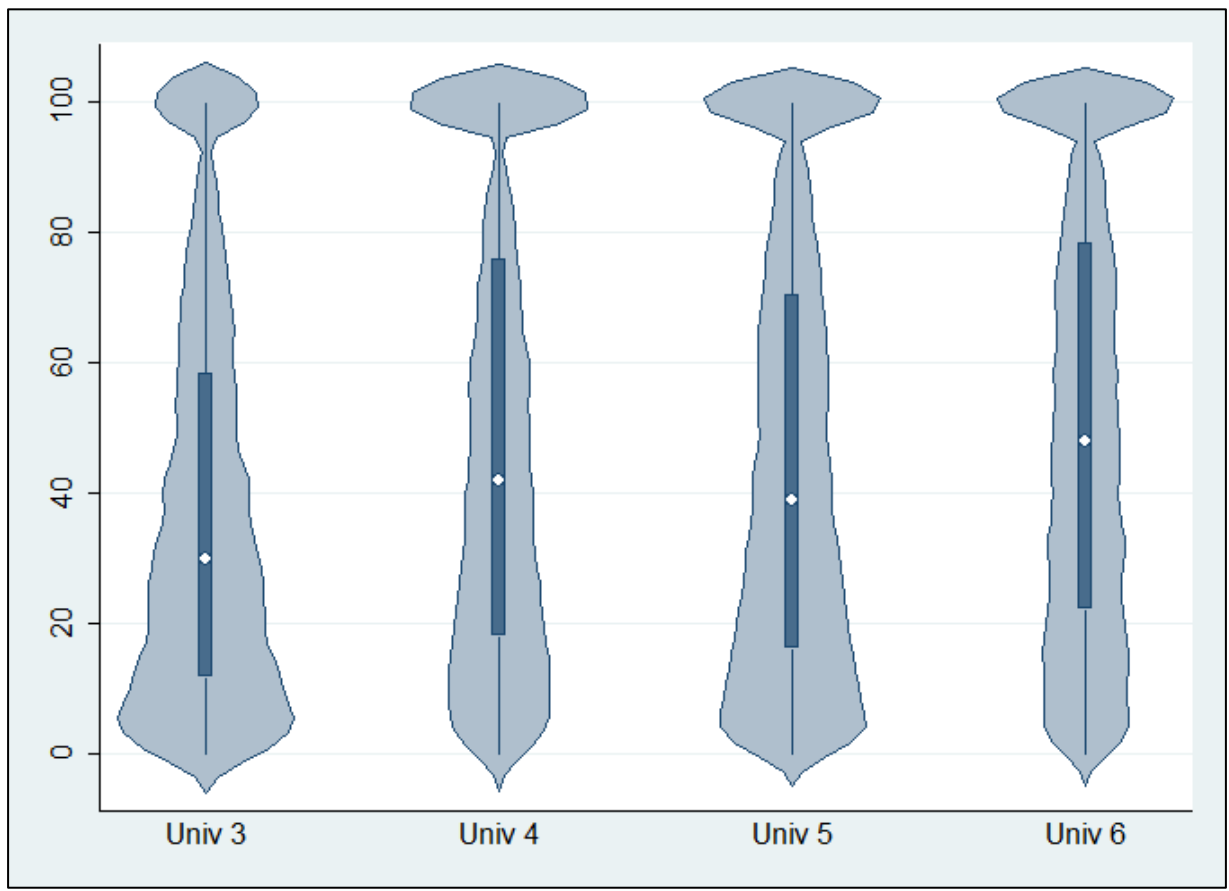

Figure 1. Distributions of inverted percentiles (i.e., high values mean low citation impact) for four universities. The violin plots show a marker for the median of the percentiles, a box indicating the interquartile range, and spikes extending to the upper- and lower-adjacent values. Furthermore, plots of the estimated kernel density (Hintze & Nelson, 1998) are overlaid.

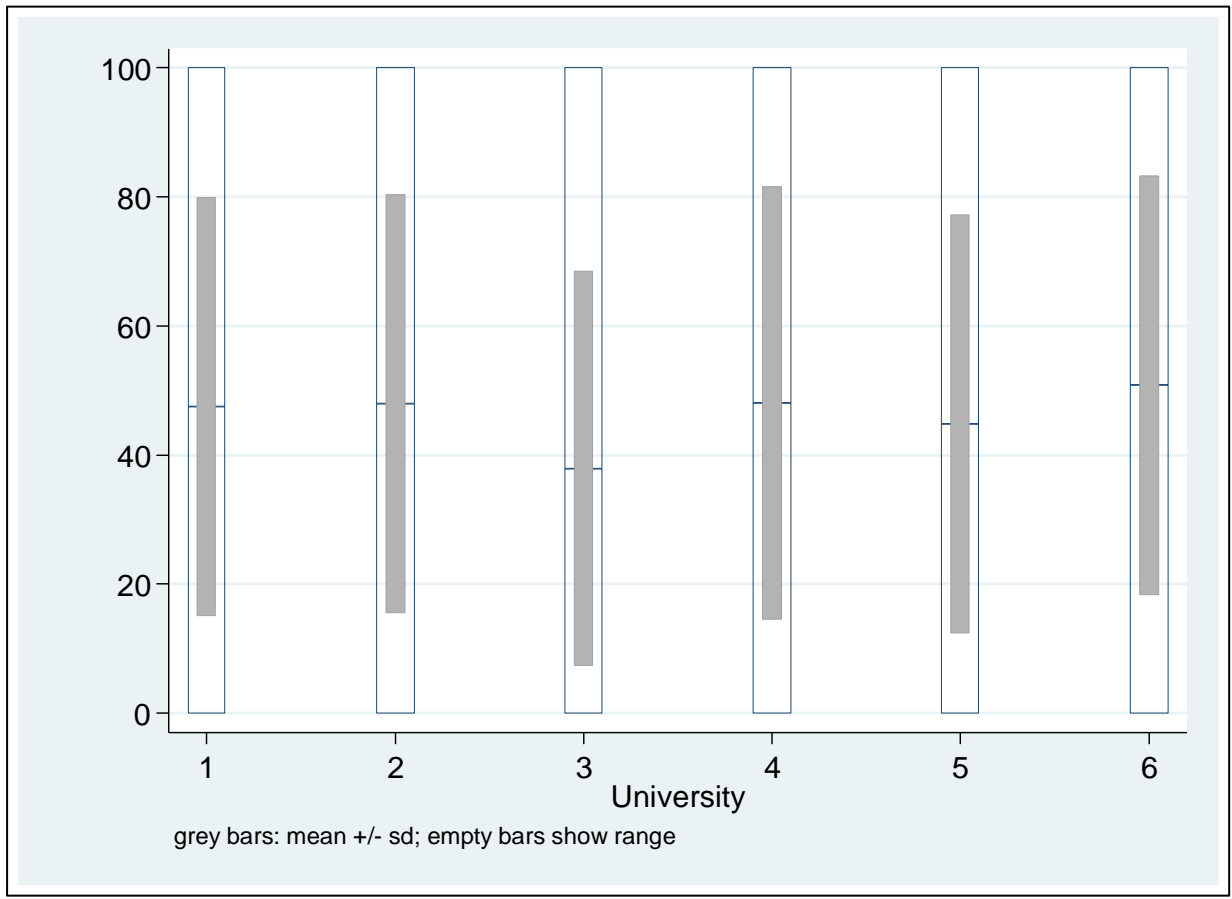

Figure 2. Inverted percentile arithmetic means, standard deviations and ranges for six universities

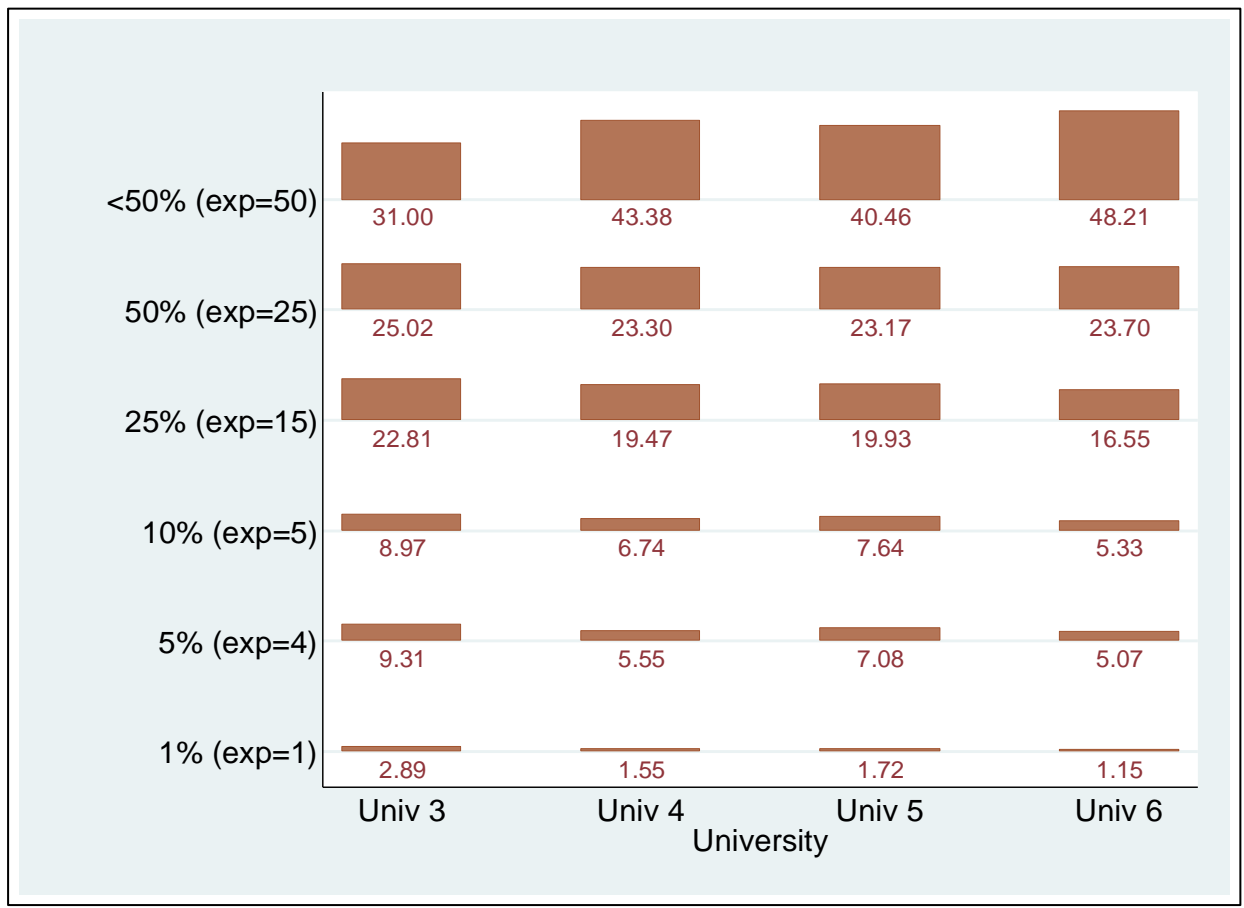

Figure 3. Differences between the universities measured by PR(6). The figure shows the percentage of a university's publications within the different classes using the bar height and as indicated below the bar. The labels of PR(6) show the percentage (exp) in brackets, which can be expected for a bar (cell) if the publications are randomly selected.